\documentclass[prc,twocolumn,nofootinbib,superscriptaddress,showpacs]{revtex4}

\usepackage{amsmath}
\usepackage{amssymb}
\usepackage{amsfonts}
\usepackage{graphicx}
\usepackage{tabularx}
\usepackage{multirow}
\usepackage{lscape}
\usepackage{rotating}
\usepackage{pstricks}
\usepackage{feynmf}
\usepackage{feynmf}

\begin{document}

\title{A new three-ranged Gogny interaction in touch with pion exchange}

\author{L. Batail}
\email{lysandra.batail@ulb.be}
\affiliation{Institute of Astronomy and Astrophysics,
Universit\'e Libre de Bruxelles, CP 226,
Boulevard du Triomphe, B-1050 Brussels, Belgium}
\affiliation{Universit\'e de Lyon, F-69003 Lyon, France; Universit\'e Lyon 1,
             43 Bd. du 11 Novembre 1918, F-69622 Villeurbanne cedex, France\\
             CNRS-IN2P3, UMR 5822, Institut de Physique Nucl{\'e}aire de Lyon}
\affiliation{CEA, DAM, DIF, F-91680 Arpajon, France}
\author{D. Davesne}
\email{davesne@ipnl.in2p3.fr}
\affiliation{Universit\'e de Lyon, F-69003 Lyon, France; Universit\'e Lyon 1,
             43 Bd. du 11 Novembre 1918, F-69622 Villeurbanne cedex, France\\
             CNRS-IN2P3, UMR 5822, Institut de Physique Nucl{\'e}aire de Lyon}
\author{S. P\'eru}
\email{sophie.peru-desenfants@cea.fr}
\affiliation{CEA, DAM, DIF, F-91680 Arpajon, France}
\affiliation{Universit\'e Paris-Saclay, CEA, LMCE, 91680 Bruy\`eres-le-Ch\^atel, France}

\author{P. Becker}
\email{becker@ipnl.in2p3.fr}
\affiliation{Universit\'e de Lyon, F-69003 Lyon, France; Universit\'e Lyon 1,
             43 Bd. du 11 Novembre 1918, F-69622 Villeurbanne cedex, France\\
             CNRS-IN2P3, UMR 5822, Institut de Physique Nucl{\'e}aire de Lyon}
\affiliation{Department of Physics, University of York, Heslington, York, Y010 5DD, United Kingdom}  
\author{A. Pastore}
\email{alessandro.pastore@cea.fr}
\affiliation{Department of Physics, University of York, Heslington, York, Y010 5DD, United Kingdom}
\affiliation{ CEA, DES, IRESNE, DER, SPRC, F-13108 Saint Paul Lez Durance, France}

\author{J. Navarro}
\email{navarro@ific.uv.es}
\affiliation{IFIC (CSIC-Universidad de Valencia), Apartado Postal 22085, E-46.071-Valencia, Spain}


\begin{abstract}
We present a new  Gogny-type finite-range effective interaction including a third Gaussian in the central term. Based on simple arguments valid for an arbitrary radial form factor, the three ranges are fixed from physical grounds, relating them to one-boson exchange interactions. Moreover, some parameters of the longest range are fixed through the G-matrix elements of the One Pion Exchange Potential. On top of giving a fairly good description of atomic nuclei properties, the resulting interaction leads to a remarkable improvement of some infinite matter properties that are relevant for astrophysical calculations.
\end{abstract}


 
\date{\today}


\maketitle


\section{Introduction} 
Nuclear Energy Density Functional (NEDF)~\cite{ben03} is the tool of choice to describe with remarkable accuracy several nuclear observables at very low computational cost~\cite{ben06,gor09,erl12}.
A convenient strategy to build well defined functionals, \emph{i.e.} free from pathologies such as self interaction~\cite{erl10,dre12,Cha10}, divergences at beyond mean-field level due to missing terms in the pairing channel~\cite{ang01} or violation of Pauli principle~\cite{dug09}, is to derive them from pseudo-potentials~\cite{rai11} containing ideally both two and three-body terms~\cite{sad13}. For more details, we refer to the discussion provided in Ref.~\cite{ben17}.

The most common family of  non-relativistic functional generator is based on zero-range Skyrme effective interaction~\cite{sky58,vau72}.  Despite its simple form, the Skyrme  functional is able to reproduce a large variety of nuclear observables such as masses~\cite{gor09}, low lying excited states of atomic nuclei~\cite{ben05} and a variety of infinite matter properties relevant for the physics of neutron stars~\cite{dav16aa}. However, as stated by the UNEDF collaboration~\cite{kor14}, the standard Skyrme energy density functional has reached its limits and significant modifications  are needed to improve the description of nuclear observables. Moreover, the use of  a Dirac delta form factor for the functional generator has the additional inconvenient of requiring a cut-off scheme to avoid ultraviolet divergences when going beyond simple mean-field calculations~\cite{bul02,mog10}.

A finite-range form factor avoids these inconveniences. The use of a Yukawa functional generator seems to be quite an obvious choice. The M3Y effective interaction~\cite{ber77}, consisting of three Yukawa terms, with their ranges fixed by exchanged boson mass, was introduced with some success to study nuclear reactions. Such an interaction was generalized by Nakada~\cite{nak03} for calculating atomic nuclei and nuclear matter properties. However, despite the strong physical motivation, there is still a long way before a Yukawa potential could be realistically usable for large scale NEDF calculations~\cite{nak06}. By replacing the Yukawa with a Gaussian, as in the Gogny interaction~\cite{dec80}, it is possible to optimize the computational cost and thus make possible large scale calculations of nuclear properties using mean field equations~\cite{dec80,per04,gor09d1m,gor16,pil17}. 

The standard Gogny interaction is composed of a central term written as the sum of two Gaussian terms $\exp{(-r^2/\mu_G^2)}$, with the four exchange operators ($1, P_{\sigma}, P_{\tau}, P_{\sigma} P_{\tau}$), plus  zero-range density-dependent and spin-orbit terms similar to those of the standard Skyrme interaction~\cite{cha97}. Within the scientific literature, only few parametrizations of the Gogny interaction exist \cite{sel14}. Most of them have been developed to accurately describe properties of atomic nuclei~\cite{gor09d1m}, but they do tend to provide very poor description of infinite matter properties, and as a consequence these interactions can not be used in the context  of astrophysical calculations. As shown in Refs.~\cite{dav16,dav17}, the standard form of the Gogny interaction does not contain enough degrees of freedom to reproduce additional constraints related to infinite matter results, as obtained from microscopic calculations based on realistic interactions.  
However, the need of a unified description of both atomic nuclei and dense nuclear star matter is very important in order to be able to maximize the information one can extract from experimental observation of both atomic nuclei and massive neutron stars~\cite{dou01,cha08,fat18}. The latter  being the most isospin asymmetric system we can probe experimentally, it provides  critical information in order to constrain the nuclear interaction. 

\section{Why a third range?}
Most of the existing Gogny interactions up to date~\cite{sel14} contain only two ranges in their central term: a \emph{short} and a \emph{long} one. As discussed in Ref.~\cite{dav17} this form does not allow to properly describe some important properties of the nuclear medium. As a consequence, Gogny interactions are not used for astrophysical calculations. Moreover, a third (long) range, as explained in Ref.~\cite{ber77}, may also improve Gogny interaction features in the context of nuclear reactions.
Among the various possible extensions of the Gogny interaction such as tensor~\cite{gra13} or finite-range density-dependent terms~\cite{cha15}, the simplest and easiest one is to modify the central term by adding a third range. This choice is mainly motivated by the possibility of implementing it into existing NEDF solvers with very minor modifications. The preliminary results presented in Ref~\cite{dav17} showed a remarkable improvement of the infinite matter properties provided by Brueckner-Hartree-Fock (BHF) calculations with realistic interactions~\cite{bal97,bal14}. However, a full refit of the parameters is mandatory to validate or not such an idea and this is one of the goals of the current work. 

\subsection{Physical determination of Gaussian ranges} 
In a full effective field theory framework, the range appearing in the radial form factor should not play any role~\cite{ben17}. Given the presence of an explicit density-dependent term however, this statement is no longer valid for the Gogny interaction, and the range of the regulator needs to be selected carefully.
Contrary to the Yukawa case, the physical interpretation of the range of a Gaussian form factor is not obvious. In that case, a full fit of the ranges is complicated and time consuming, and the ranges have been manually adjusted from the start guided by the idea that a phenomenological transcription of the nuclear force should have a short range repulsive and a long range attractive term. For instance, the range values were fixed to $0.7, 1.2$~fm and $0.5, 1.0$~fm for interactions D1~\cite{dec80} and D1M~\cite{gor09d1m}, respectively.
As shown in Ref.~\cite{dav17}, it is not easy to identify a good set of observables that may constraint the third range, thus making a direct fit of the range a complicated task.
In this article, we suggest to identify the ranges of an effective interaction using an approach based on the microscopic considerations of the ratio of the exchange (Fock) over the direct (Hartree) contributions to the nucleon self-energy. Once these values have been fixed, we show that it is then possible to perform a complete fit of the parameters including relevant properties of infinite matter and atomic nuclei, but also conserving some relevant asymptotic properties of the long range part  \cite{ber77}.

As far as a Yukawa potential is concerned, expressed as $e^{-\mu_Yr}/\mu_Yr$, the physical interpretation of the range is clear: $\mu_Y^{-1}$ is deduced from the exchanged meson mass such as $\mu_Y= m c/\hbar$. 
 For a given mass or range, the ratio Fock/Hartree 
must actually reflect the microscopic properties of the underlying nuclear interaction. For a Yukawa form factor in infinite nuclear matter, such a ratio reads  
\begin{equation}\label{eq:yuk}
R_{Y}=\frac{3}{16x_Y^4}\left[  2x^2_Y - 2 x_Y \arctan 2x_Y +\text{arctanh} \frac{2x_Y^2}{1+2x_Y^2} \right],
\end{equation}
\noindent where  $x_Y$ = $k_F / \mu_Y$ and $k_F$ is the Fermi momentum.
Notice that $R_Y \to 1/4$ when $x_Y \to 0$, whereas $R_Y \to 0$ as $x_Y \to \infty$. 
This illustrates well that for a short-range potential the exchange and direct terms give the same contribution (up to the degeneracy factor due to the sum over all possible states in the Hartree term, $(2S+1)(2T+1)=4$ in our case), while for a long-range potential the exchange term becomes negligible.

For a Gaussian form factor in infinite nuclear matter one gets 
\begin{eqnarray}\label{eq:gog}
R_G=\frac{3}{4x_G^4}\left[  -2 + 2 \, e^{-x_G^2}+\sqrt{\pi} \, x_G \, \text{Erf}(x_G) \right]\;,
\end{eqnarray}
\noindent where $x_G$ = $k_F \; \mu_G$. 

\begin{figure}[!h]
\begin{center}
\includegraphics[width=0.52\textwidth,angle=0]{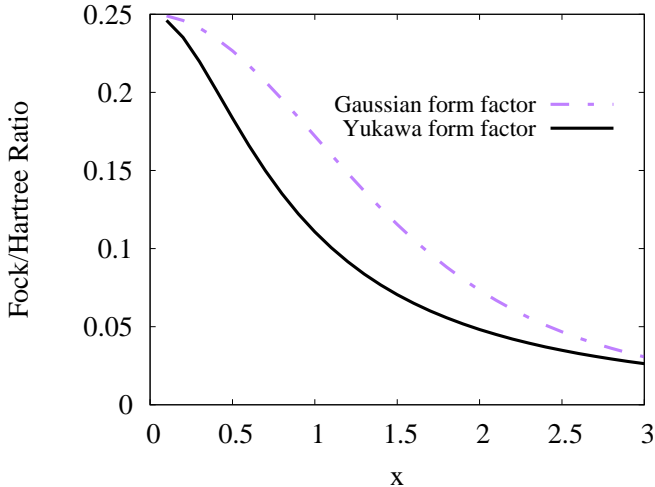}
\end{center}
\caption{Evolution of the ratios $R_{Y,G}$ as a function of the dimensionless parameter $x$. Solid line stands for Eq.~(\ref{eq:yuk}) for a Yukawa form factor, while dashed line stands for Eq.~(\ref{eq:gog}) obtained with a Gaussian form factor. See text for details.}
\label{portee}
\end{figure}

The ratios $R_{Y}$  and $R_G$ are plotted in Fig.~\ref{portee} as a function of the dimensionless parameter $x$. We observe that the behavior at the limits $x \to 0$ and $x \to \infty$ is quite general and not related to the particular choice of the form factor. 

Now, we fix the range $\mu_G$ of a Gaussian form factor so that the Fock/Hartree ratio $R_G$ is equal to $R_Y$ for a given exchanged meson at the same Fermi momentum. In that way, Gaussian ranges are related to exchanged mesons in the Yukawa potential. 
To be specific, we consider as a starting point the work of Ref.~\cite{ber77} where the M3Y nuclear interaction has been fitted using three Yukawa functions with ranges taken as the Compton wavelength of $\pi$, $\sigma$ and $\rho$ mesons with masses values of roughly 770, 490 and 140 MeV. These masses (or ranges) give us a solid pathway to build the effective interaction: the long-range scale is for instance governed by the one-pion exchange process while the short-range part is governed by two-pion exchange (mocked by the $\rho$ meson). 
As compared to a direct fit over some selected observables, this approach provides us with a physical insight of the roles played by the different ranges. 

For the three masses under consideration and with $k_F = 1.33$ fm$^{-1}$ for the Fermi momentum, we obtain the following Fock/Hartree ratios
\begin{eqnarray}
R_Y(\mu_Y^{-1} = 0.256 \text{ fm} )& = & 0.214\;,\\
R_Y(\mu_Y^{-1} = 0.402 \text{ fm} )& = & 0.178\;,\\
R_Y(\mu_Y^{-1} = 1.407 \text{ fm} )& = & 0.052\;.
\end{eqnarray}
This means for instance that at saturation density the Fock term is about 19 times smaller than the Hartree term for the pion.
Imposing now that the Gaussian form factor reproduces the same physics as the Yukawa one (\emph{i.e.} the same numerical values of the ratio $R_Y$), we obtain
\begin{equation}
\mu_1=0.475,\; \mu_2=0.716, \;\mu_3=1.78\mbox{  fm}.
\label{theranges}
\end{equation}
It is worth mentioning that the shorter range parameters of interactions D1 and D1M are respectively close to $\mu_2$ and $\mu_1$ obtained here, while their longer range values lie between $\mu_2$ and $\mu_3$ for both interactions. The results obtained in this way depend on the choice of $k_F$: doing the same exercise at half-saturation instead of at saturation density, we obtain $\mu_1=0.483,\; \mu_2=0.740, \;\mu_3=1.93\mbox{  fm}$. 

\section{Fitting the interaction}
Employing a fitting protocol similar to the one adopted in Ref.~\cite{cha97}, we have obtained a new Gogny-like interaction, named D3G3, composed of three Gaussians for the central term plus density-dependent and spin-orbit terms analogous to the usual Skyrme-like form~\cite{cha97}.  We leave for a future work a more thorough investigation concerning the incorporation of other extra terms, like the tensor one to improve for instance the shell structure~\cite{gra13}.

\begin{figure}
\begin{center}
\includegraphics[width=0.46\textwidth,angle=0]{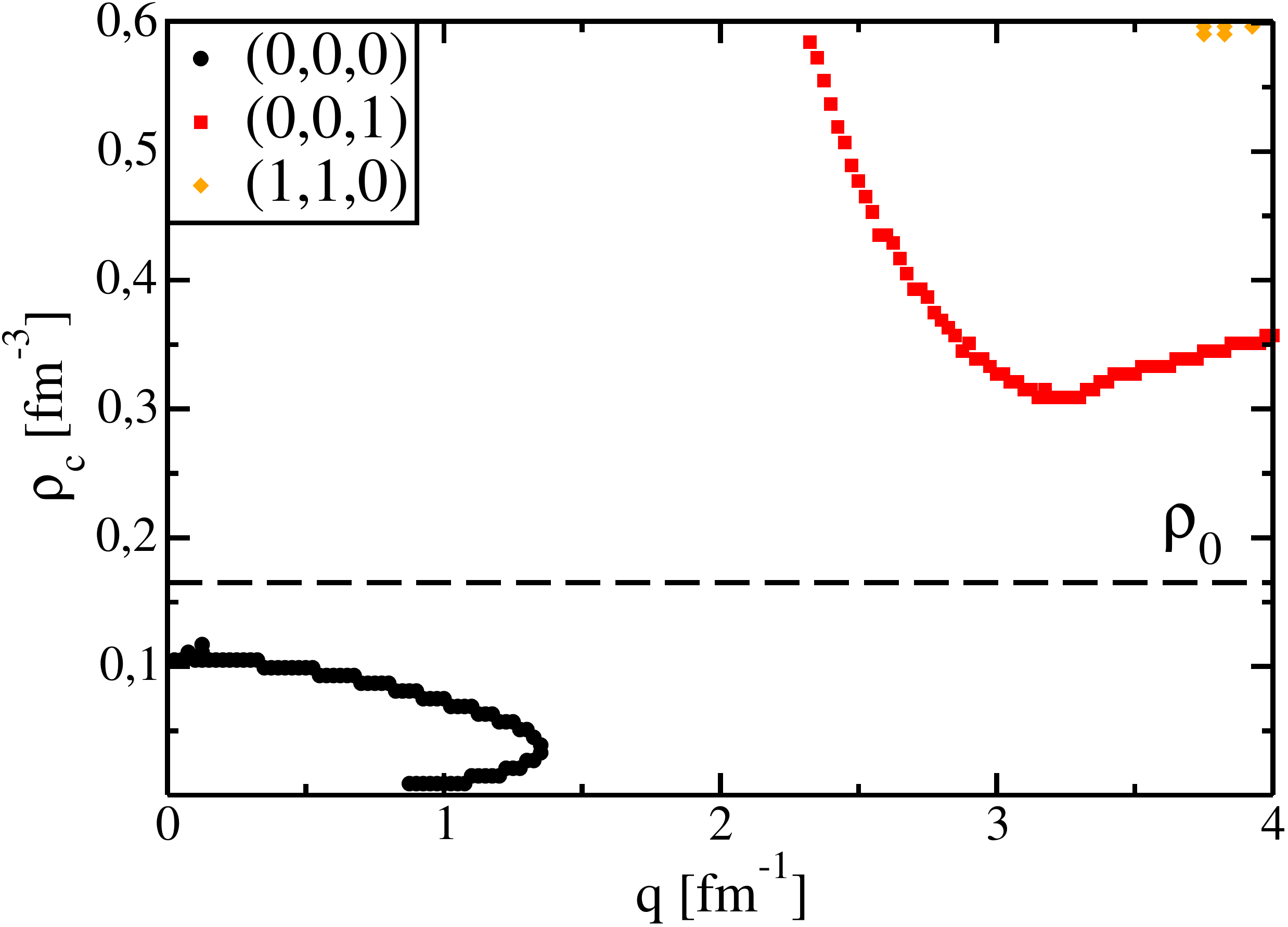}
\end{center}
\caption{Critical densities in as a function of the transferred momentum for the D3G3 parametrisation in the various spin, isospin channels. The horizontal dashed line represent the value of saturation density $\rho_0$.  See text for details.}
\label{crit}
\end{figure}

To keep close with the OPEP physics, we followed the procedure highlighted in Ref.~\cite{nak03} and fixed 3 out of the 4 parameters $\{W_3, B_3, H_3, M_3\}$ related to the long range term, as originally suggested in Ref.~\cite{ber77}. As already mentioned, such a choice provides us with good asymptotic properties, of importance for nuclear reactions. We were consequently left with only one additionally free parameter associated to the third Gaussian compared to a standard Gogny interaction~\cite{dec80}. It is important to stress here that in the fit we have performed, the ranges were actually adjusted. Therefore, $\mu_{1,2,3}$ values numerically determined above (Eq.~\ref{theranges}) must only be considered as a starting point of the minimisation process. 
We let them evolve afterwards, on an equal footing with the other free parameters, in order to minimize the penalty function $\chi^2$. What we found in the end is that the ranges are less than 10\% different from first values, thus showing our initial guess is well motivated. The full set of parameters is reported in Tab.~\ref{tab:par}.
In addition, by means of the linear response theory~\cite{dav21}, we have calculated the position of the critical densities, $\rho_c$, at which instabilities appear in symmetric nuclear matter (SNM)~\cite{sch19book} as a function of the transferred momentum $q$. The results are given in Fig.~\ref{crit} for the various spin (S), spin-projection (M) and isospin (I) channels.
In the scalar/isoscalar channel (S=0,M=0,I=0), we observe the appearance of the spinodal instability~\cite{duc07} that identifies the boundaries of the gas/cluster phase transition in SNM.
In the scalar/isovector channel (S=0,M=0,I=0), we observe another instability that relates to a separation of neutron and protons. The lowest critical densities are well above two times saturation density.
According to the stability criterion discussed in Ref.~\cite{hel13}, we can conclude that these instabilities do not affect calculations of properties of finite nuclei for D3G3. We refer the reader to the review articles~\cite{report,dav21} for a more detailed discussion on this topic.

Being free from these instabilities is a basic requirement for an effective interaction so as to obtain meaningful results in finite nuclei. As an example, one can mention D1N~\cite{cha08} and D1M*~\cite{gon18}  as displaying isovector spurious finite size instabilities that prevent a simple Hartree-Fock code in coordinate space~\cite{mar19,gon21} from converging. Both interactions hence require a careful selection of the basis properties, for example using an harmonic oscillator basis in order to handle such a divergence~\cite{gon21} and be able to obtain a fully converged result at the mean-field level. Unfortunately, finite-size instabilities may also impact the position of low-lying excited states~\cite{pas15} and in this case no clear procedure on how to handle such instabilities has been suggested yet, thus making interactions affected by such pathology less appealing for large scale calculations of atomic nuclei global properties. 

\begin{table}[!h]
    \centering
   \begin{tabular}{c| c |c| c| c |c}
    \hline
 $i$   & $W_i$ (MeV)& $B_i$ (MeV)& $H_i$ (MeV)& $M_i$ (MeV) & $\mu_i$ (fm)\\
    \hline
1   &  -7543.80& 13485.57&-14708.99&6669.46&0.470\\
2 & 590.47&-1751.40&1582.84&-909.26&0.749\\
3&  4.63& -9.25& 9.25& -18.50 & 1.967\\
    \hline
    \multicolumn{6}{c}{$t_3$=1400 (MeVfm$^{3(\alpha+1)}$)  $x_3=1$  $\alpha=1/3$ $W_0=115.14$ (MeVfm$^5$) }\\
    \hline
\end{tabular}
\caption{Parameters of central, density dependent and spin-orbit terms of the D3G3 interaction. }
\label{tab:par}
\end{table}

\subsection{Infinite nuclear matter}
Following a previous work on the SNM equation of state decomposition in partial waves~\cite{dav16}, we used as constraints the global equation of state (EOS) and its decomposition in the four spin $S$-isospin $T$ channels.
It is worth mentioning that these channels are not observables but can be obtained using different \emph{ab-initio} methods. Hence, results may differ from one to another, especially beyond saturation density ~\cite{Dav15} although the general trends remain similar. For such a reason, we do not give a strong weight to these quantities in the fit, but we use them only to provide a reasonable repartition of the strength of the interaction in the various channels. We have decided as a consequence to limit our fit to values of Fermi momentum $k_F\le1.33$ fm$^{-1}$  constraining saturation properties to the values currently adopted in the literature~\cite{dut12}.

\begin{figure}[!h]
\begin{center}
\includegraphics[width=0.52\textwidth,angle=0]{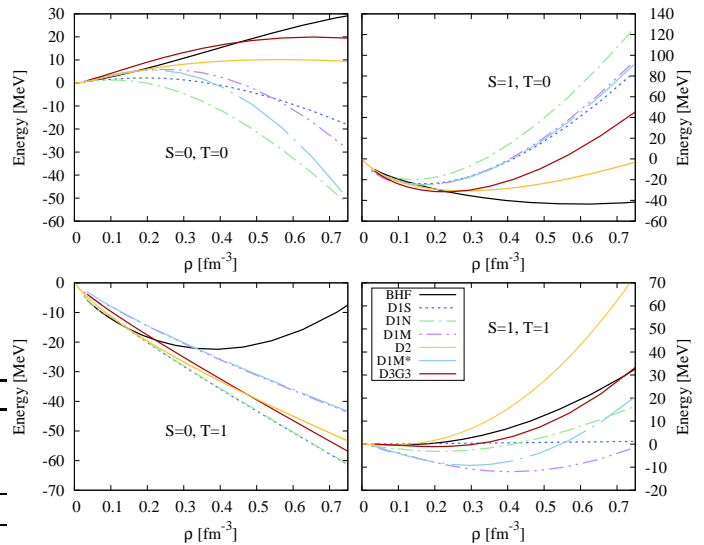}
\end{center}
\caption{Spin-isospin decomposition of the EOS for various Gogny interactions and the BHF calculations. See text for details.}
\label{ST:channel}
\end{figure}

In Fig.~\ref{ST:channel}, we show the behavior of  D3G3, as well as several standard Gogny interactions, in the different $(S,T)$ channels along with results obtained within the Brueckner-Hartree-Fock (BHF) approximation~\cite{bal97,bal14}. The choice of BHF with respect to any other available microscopic method is mainly motivated by the variety of available results and allows direct comparison with previous works~\cite{dav16}. Besides, as highlighted in Ref.~\cite{Dav15}, different many-body methods provide us with compatible results concerning the four $(S,T)$ channel up to saturation density.
We notice that D3G3 reproduces quite nicely all the channels apart from the low density part of the $S=0,T=0$, contrary to the other Gogny interactions. The $S=0,T=1$ channel being sensitive to pairing properties, we ensured we were able to have the right trend at low density. This was a good starting point to provide an interaction capable to show a reasonable description of properties of open shell nuclei. Also, both the low density and general behaviour of the $S=1,T=1$ channel fit nicely BHF results. This better reproduction is due to a gain in flexibility with respect to the standard structure of the Gogny interaction based on 2 Gaussians. 
The only exception being D2~\cite{cha15} which has a non-standard form since it contains a finite range density dependent term. Thanks to this term, the D2 interaction has good infinite matter properties as compared to other Gogny interactions~\cite{sel14}, but at a remarkably increased level of complexity in order to be used in a numerical solver to treat finite nuclei especially for beyond mean-field calculations as GCM~\cite{rob18} and QRPA~\cite{per04}.
However, one notices that none of the parameter sets displayed in Fig.~\ref{ST:channel} is capable to reproduce the curvature of the $S=0,T=1$ channel. This may be due to some limitations remaining in the force and leaves room for further development.
At the end of our fit, we obtained a saturation density $\rho_0=0.165$  fm$^{-3}$  with an energy per particle, $E/A=-16.05$ MeV. The effective mass at saturation takes the value $m^*/m=0.68$ and the nuclear incompressibility $K_\infty=227$ MeV.
These numbers imply that D3G3 is capable to provide a good description of bulk properties of nuclear matter. 

\begin{figure}[!h]
\begin{center}
\includegraphics[width=0.5\textwidth,angle=0]{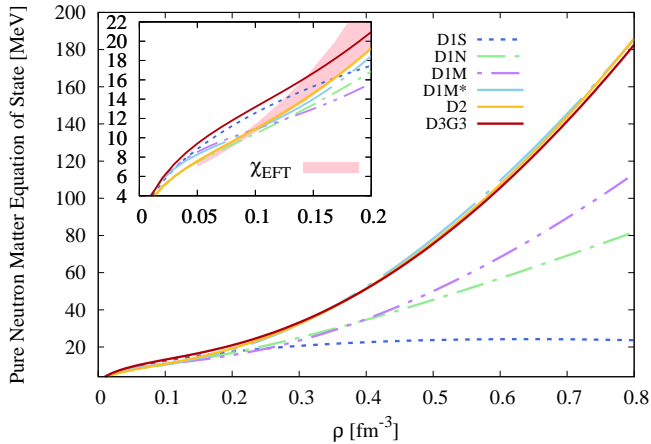}
\end{center}
\caption{Equation of state in PNM for various Gogny interactions. The shaded area correspond to the EOS obtained using many-body calculations based on chiral two-, three-, and four-nucleon interactions~\cite{dri19}.}
\label{PNM}
\end{figure}

In order to perform astrophysical calculations, another critical quantity to be evaluated is the equation of state of pure neutron matter (PNM). Our results are displayed in Fig.~\ref{PNM}. 

Thanks to the additional flexibility of the third Gaussian, we have been able to obtain a stiffer EOS which is compatible with most of current results obtained with  \emph{ab-initio} methods. As an illustration, we report on the same figure as a shaded area the results obtained using many-body calculations based on chiral effective field theory ($\chi$EFT) taken from Ref.~\cite{dri19}. It is worth mentioning that D1M* and D2 also provide a very stiff EOS suitable for astrophysical calculations.
By using the EOS of D3G3 to solve the Tolman-Oppenheimer-Volkoff equations, we have studied the macroscopic properties of a massive neutron star (NS)~\cite{tho00}. As done in Ref~\cite{sel14} we assume the NS made only by neutrons, but this approximation has small impact on our conclusions. We have verified that the heaviest NS sustained by such an EOS has a mass of $2.04$ $M_\odot$. This value is in good agreement with the most massive NS observed up to date~\cite{ant13}.
More recently, the use of gravitational waves (GW) has offered the opportunity to obtain additional constraints on the radius of a NS with $1.4M\odot$ as observed in the event  GW170817~\cite{abb17}.
In particular, we obtained a radius of $R_{1.4M\odot}=11.06$ km, again in agreement with the interval deduced in  Ref~\cite{ant13,ann18} and deduced from GW analysis.
Similar results can be obtained using D1M* and D2~\cite{gon18}.
It is interesting to observe that according to Ref.\cite{sel14}, two other Gogny interaction provide NS properties compatible with current observations: namely D1AS~\cite{ono03} and D1P~\cite{far99}. Both interactions contains additional density dependent terms that allow for a stiffer EoS in PNM.

\begin{figure}[!h]
\begin{center}
\includegraphics[width=0.5\textwidth,angle=0]{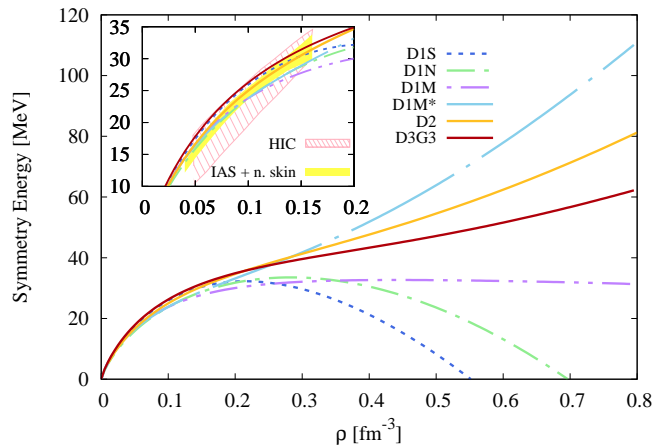}
\end{center}
\caption{Density dependence of the symmetry energy for various Gogny interactions. The shaded area in the insert correspond to the constraints obtained via HIC~\cite{dan14}  and isobaric analog states and neutron skin analysis ~\cite{tsa09}.}
\label{Esym}
\end{figure}

Finally, in Fig.~\ref{Esym}, we show the evolution of the symmetry energy $E_{sym}$ as a function of the density of the infinite medium. Some Gogny interactions as D1S and D1N lead to a collapse of $E_{sym}$ at high density. This is related to a nonphysical collapse of the EOS in PNM. At low density some constraints coming from heavy ion collisions \cite{dan14} and the analysis of isobaric analog states plus neutron skin ~\cite{tsa09} are available: D3G3 is compatible with them.
The behavior of D3G3 at high density differs from the one of D1M* and D2 and this may have an impact on the direct URCA process in NS cooling process~\cite{hae95}, but a detailed analysis of such a case is beyond the scope of this work.

\begin{figure}[!h]
\begin{center}
\includegraphics[width=0.5\textwidth,angle=0]{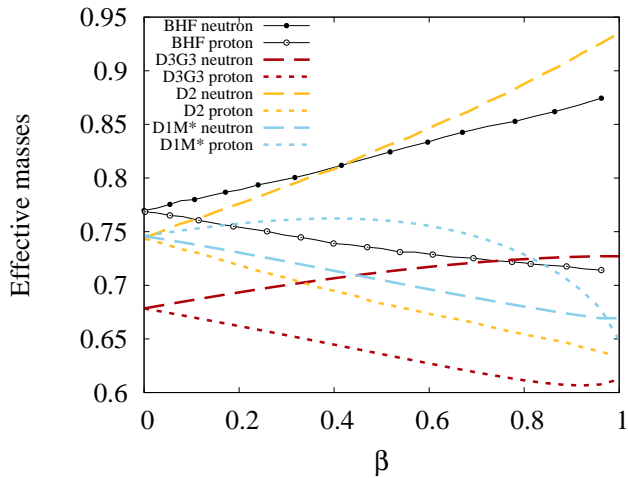}
\end{center}
\caption{Evolution of the neutron (dashed-line) and proton (dotted-line) effective mass for some relevant Gogny parametrisation as a function of the asymmetry parameter $\beta$. On the same figure we report the BHF results (dots). See text for details. }
\label{mstar}
\end{figure}

In Fig.~\ref{mstar} we show the evolution of the effective masses as a function of the asymmetry parameter $\beta=(\rho_n-\rho_p)/\rho$, where $\rho_{n(p)}$ is the neutron (proton) density.
We compare our results obtained with the BHF ones of Ref.~\cite{bal14}.
The three Gogny interactions analysed here, namely D2, D1M* and D3G3, have the correct isovector splitting compared to the \emph{ab-initio} results, although the effective mass of D3G3 is slightly lower than the others. The mass difference between neutrons and protons is however of the right order of magnitude. This is not the case for example in D1M* where we observe almost an equal proton/neutron mass in PNM.

\subsection{Atomic nuclei} 
As a preliminary \textit{in situ} analysis, we have calculated the binding energy of $\sim$2000 nuclei with prescriptions very similar to the D1M ones ~\cite{gor09d1m} and obtained a global root mean square deviation (RMS) of $\sigma_{\text{RMS}} \simeq$ 7 MeV. We recall that since our fit included the masses of only three double-magic nuclei: $^{48}$Ca, $^{56}$Ni and $^{208}$Pb, it is not surprising to obtain such a large value of RMS. A fit over this large an amount of nuclei goes nevertheless beyond the scope of this article and is left for a future study. It is however interesting to note that this result is comparable with the one obtained using D2~\cite{pil17} but shows that using only few nuclear masses into a fit is not sufficient to obtain a lower RMS. A quick inspection of the shell structure of Sn isotopes calculated using D3G3 shows that one can recover a shell structure comparable with the one obtained with other Gogny interaction, although the resulting pairing gaps appear to be small compared to available experimental data.
To date, only the D1M parametrisation fitted to reproduce binding energies beyond mean-field approaches can compete with phenomenological mass-formulas~\cite{sob14}. Such an extension of the global fit procedure should be possible to implement for D3G3 since only the number of ranges in the central term is increased by one.

\section{Conclusions}
We have presented a new way of understanding the ranges of a gaussian form factor by rooting them from the masses of exchanged bosons. We have applied this idea to a new family of Gogny interaction by including three effective ranges, corresponding to the masses of pion, rho and sigma mesons as it happens for more realistic interaction based on Yukawa form factors. 

By means of a simple fitting protocol, we have shown that it is possible to obtain a Gogny interaction with only one additional parameter that can be easily incorporated in existing numerical codes, free from instabilities and rather suited for astrophysical calculations. Moreover, results obtained with D3G3 are on average better than several existing Gogny parametrizations regarding infinite matter properties. 
Even if a refined work remains necessary for large-scale nuclear structure calculations, D3G3 can however be considered as a valuable starting point for future work.

\section*{Acknowledgements}
The authors would like to thank N. Pillet for useful discussions about the D2 interaction and S. Goriely for discussions about masses.


\bibliography{biblio_JN}

\end{document}